\documentstyle[12pt]{article}

\newcommand{\be}{\begin{equation}}
\newcommand{\ee}{\end{equation}}
\newcommand{\bea}{\begin{eqnarray}}
\newcommand{\eea}{\end{eqnarray}}
\newcommand{\ba}{\begin{array}}
\newcommand{\ea}{\end{array}}
\newcommand{\vs}[1]{\vspace{#1 mm}}

\newcommand{\e}{\epsilon}

\def\bbox{{\,\lower0.9pt\vbox{\hrule \hbox{\vrule height 0.2 cm
\hskip 0.2 cm \vrule height 0.2 cm}\hrule}\,}}
\newcommand{\dsl}{\pa \kern-0.5em /}

\newcommand{\shalf}{\frac{1}{2}}
\newcommand{\pa}{\partial}

\newcommand{\nn}{\nonumber\\}

\font\mybb=msbm10 at 12pt
\def\bb#1{\hbox{\mybb#1}}

\def\bE {\bb{E}}

\def\sac{\, , \qquad}
\def\mf{M5-brane }
\def\mt{M2-brane }

\def\cth{C_{(3)}}

\def\csix{C_{(6)}}

\def\gtil{\tilde{g}}

\begin{document}

\topmargin 0pt
\oddsidemargin 5mm

\renewcommand{\thefootnote}{\fnsymbol{footnote}}
\begin{titlepage}

\setcounter{page}{0}
\begin{flushright}
ECM-UB-98/06 \\
UTTG-03-98\\
hep-th/9803040
\end{flushright}

\vs{5}
\begin{center}
{\Large BRANE-INTERSECTION DYNAMICS FROM BRANES IN BRANE BACKGROUNDS}
\vs{10}

{\large
Joaquim Gomis$^1$\footnote{Permanent adress University of
Barcelona}, David Mateos$^2$, Joan Sim\'on$^2$ and 
Paul K. Townsend$^3$\footnote{On
leave from DAMTP,  University of Cambridge, U.K.}
} \\
\vs{5}
${}^1${\em Theory Group, Department of Physics,\\
University of Texas, Austin, TX 78712}\\
\vs{5}
${}^2${\em Departament ECM, Facultat de F{\'\i}sica, \\
Universitat de Barcelona and Institut de F{\'\i}sica d'Altes Energies, \\
Diagonal 647, E-08028 Barcelona, Spain}  
\vs{5}
${}^3${\em Institute for Theoretical Physics\\
University of California, Santa Barbara\\
CA 93106, USA}
\end{center}
\vs{10}
\centerline{{\bf{Abstract}}}

We derive the dynamics of M-brane intersections from the worldvolume action of
one brane in the background supergravity solution of another one. In this way
we obtain an effective action for the self-dual string boundary of
an M2-brane in an M5-brane, and show that the dynamics of the 3-brane
intersection of two M5-branes is described by a Dirac-Born-Infeld action.

\end{titlepage}
\newpage
\renewcommand{\thefootnote}{\arabic{footnote}}
\setcounter{footnote}{0} 

\section{Introduction}

Intersecting M-branes are becoming increasingly important in many aspects of
non-perturbative QFT and quantum gravity. Our interest here will be with 
1/4-supersymmetric orthogonal intersections of two M-branes. These have been 
investigated, and classified, by a variety of methods. From the perspective of
the worldvolume of one of the participating M-branes the intersection with the
other one appears as a 1/2-supersymmetric soliton-type solution of its
worldvolume field theory. A notable example is the self-dual string soliton in
the M5-brane \cite{HLWa} which can be interpreted as the boundary of an M2-brane
\cite{strom,pkt} (this is the M-theory version of the Dirac-Born-Infeld
`BIons' \cite{CM,gib} which can be interpreted as the endpoints of `fundamental'
strings on D-branes). Another example is the 3-brane soliton in the M5-brane
\cite{HLWb}, which can be interpreted as the intersection with another 
M5-brane \cite{paptown}. All of these worldvolume solitons saturate a
Bogomolnyi-type bound in terms of a central charge appearing in the worldvolume
supersymmetry algebra \cite{ggt},  and consideration of these charges suffices
to classify all possible 1/4 supersymmetric intersections \cite{bgt}. 

Here we address the issue of the effective actions describing the dynamics of 
the self-dual string and 3-brane solitons of the M5-brane. These are expected to
be $\kappa$-symmetric string and 3-brane actions in the 6-dimensional Minkowski
background provided by an infinite static planar M5-brane. The worldvolume
fields are in correspondence with the zero modes in an analysis of  fluctuations
about the worldvolume soliton solutions, and this type of analysis has been
carried out in \cite{HLWc}\cite{george}. Here we take a different approach. We consider the
worldvolume field theory of a `test' M-brane in the background spacetime of an
M5-brane. This is a justifiable approximation if the source of the `supergravity
M5-brane' is actually a large number of coincident M5-branes. The approach is
similar to one adopted in a number of recent works in which a brane of M-theory
or string theory is put into the background of a large number of parallel branes
of the same type \cite{malda,kal}. The `test' brane feels no force in this
background because it is parallel to the `source' brane. Our work exploits the
fact that there are various other embeddings of test M-branes in the same
background for which the test brane again feels no force. In fact, such
embeddings correspond precisely with the possible 1/4 supersymmetric
intersections of an M5-brane \cite{tsey}.

While an M2-brane boundary on an M5-brane appears as a worldvolume string
soliton of the M5-brane's worldvolume field theory, there is no similar
interpretation of this `intersection' from the M2-brane's point of view,
essentially because boundaries are determined by imposing boundary conditions
rather than by solving field equations. One motivation for the approach
taken here is that it circumvents this difficulty. When the M5-brane is replaced
by its supergravity solution the M2-brane actually has no boundary, it just
disappears down the infinite M5-brane `throat'. There is therefore no need to
impose boundary conditions on the M2-brane equations. Nevertheless, on scales
that are large compared to that determined by the M5-brane tension the 
supergravity solution can be replaced by an effective 5-brane source, and it 
will then appear that the M2-brane has a boundary on the M5-brane. We can
therefore study the dynamics of this boundary by considering fluctuations of the
membrane in the M5-brane background. In this way, we are able to derive an
effective action for the string boundary. The string tension is formally
infinite, but this is to be expected of an infinite membrane. By considering a
membrane stretched between two M5-branes the tension is made finite. As we
shall see, this is true even though the {\sl proper} length of the membrane
in the direction separating the M5-branes is infinite.

The 3-brane intersection of two M5-branes can be treated in the same way.  In
this case we consider the fluctuations of a test M5-brane embedded in an
appropriate way in the background of a source M5-brane. The resulting effective
action, for which the (partially gauge-fixed) fields are those of a D=6
vector supermultiplet \cite{HLWb}, is of Dirac-Born-Infeld (DBI) type. From the
work of \cite{CM,gib} it then follows that this 
3-brane has its own worldvolume
`bions' which can be interpreted as the endpoints of self-dual strings. We
therefore confirm the claim of \cite{bgt} that the 3-brane soliton is a D-brane
for the self-dual string soliton.


\section{The \mt\ ending on the \mf}

Our starting point will be the action for the supermembrane in a D=11
supergravity background \cite{m2action}. To specify the latter we must, in principle,
choose a background supervielbein $E_M{}^A$ and 3-form superspace gauge-potential
$C_{(3)}$ satisfying the on-shell superfield constraints of D=11 supergravity.
The field equations of this action are the M2 `branewave' equations. We shall
choose a purely bosonic background for which the fermion equations are
trivially solved by setting them to zero. Equivalently, we can start by
discarding the worldvolume fermions, in which case the action is 
\be
S_{M2} = - \int d^3\xi \sqrt{-\det g} + \int_W {\cal C}_{(3)}
\label{m2action}
\ee 
where $g$ is the metric induced from the spacetime 11-metric and ${\cal
C}_{(3)}$ is now the pullback of the spacetime 3-form potential to the
worldvolume $W$ (with coordinates $\xi^I$, $I=0,1,2$). We shall take the
background to be that of the M5-brane solution.  This is a purely bosonic
background with 11-metric and 4-form field strength
$F=dC_{(3)}$ given by \cite{guven}
\bea
ds^2_{(11)} &=& U^{-1/3} \eta_{\mu\nu}dY^\mu dY^\nu + U^{2/3}
d{\bf X}\cdot  d{\bf X} \nn
F_{mnpq} &=& \epsilon_{mnpqr}\partial_r U
\eea
where $\eta_{\mu\nu}$ is the metric on the 6-dimensional Minkowski space with
$Y$ coordinates, and $U$ is a harmonic function on the transverse 5-space
$\bE^5$ with cartesian coordinates $X^m$ and euclidean metric $d{\bf X}\cdot
d{\bf X}$. To begin with we choose
\be
U=1+\frac{q}{r^3}\, ,  
\ee
where $r$ is the radial distance from the origin in $\bE^5$. 

We shall first seek a static solution of the membrane field equations in this
background that can be interpreted as the linear orthogonal intersection of an
M2-brane with an M5-brane. Setting $\xi^I =(\sigma^i,\rho); (i=0,1)$, we are 
thus led to make the partial gauge choice\footnote{The choice
$X^1=f(\rho)$ for any monotonic function $f$ would be equally good, but the
range of the membrane coordinate $\rho$ will depend on the choice, as discussed
below. Here we make the simplest choice.}
\be\label{pgc}
X^1 = \rho\, 
\ee
combined with the ansatz
\be
Y^{0}=\sigma^0 \sac Y^{1}=\sigma^1\, ,
\label{m2m5.vac.sol}
\ee
with all other worldvolume fields vanishing. It is straightforward to verify 
that this membrane configuration solves the branewave equations. The solution
represents a membrane that `disappears' down the infinite M5-brane throat at
${\bf X}={\bf 0}$. On the surface $X^2=X^3=X^4=X^5=0$, the proper distance to
$X^1=0$, i.e. $\rho=0$, is infinite. This means that $\rho=0$ does not
correspond to any points of the membrane; the coordinate $\rho$ therefore takes
values in the {\sl open} interval $(0,\infty)$. 

Although the membrane has no boundary it will appear to end on the M5-brane on
length scales for which the M5-brane background can be replaced by an effective
M5-brane source. It should therefore be possible to determine the dynamics of
this effective membrane boundary from the dynamics of the membrane itself. To do
so we must consider the (not-necessarily small) fluctuations about the above
solution of the branewave equations. To proceed, we restrict the oscillations of 
the \mt\ to those obeying the  following conditions:
\be
Y^{\mu} = Y^{\mu}(\sigma) \sac  X^{1}=\rho \sac 
X^{2}=X^{3}=X^{4}=X^{5}=0\, .
\label{m2.complete.ansatz}
\ee
These restrictions force the membrane oscillations to be uniform in the $X^1$
direction. On sufficiently large length scales this will be interpretable as a
membrane oscillating rigidly with its boundary in an M5-brane, the boundary
oscillations being unrestricted. The restrictions (\ref{m2.complete.ansatz}) 
also constitute a consistent truncation. In particular, the branewave equations
for the $X$ fields are automatically satisfied. To verify this it is crucial to
observe that the pull-back 3-form ${\cal C}_{(3)}$ vanishes for worldvolume
fields satisfying the above conditions\footnote{And hence for our `vacuum'
solution of these equations; this fact was implicitly used earlier.}. Because
the M5-brane supergravity solution is such that $F$ is a 4-form on the 5-space
with coordinates $X$, one can choose the 3-form potential $C_{(3)}$ such that it
too is a form on this 5-space. It follows that the pullback of $C_{(3)}$ to the
worldvolume involves derivatives of at least three different $X$ coordinates,
only one of which can be non-zero for fields satisfying the ansatz 
(\ref{m2.complete.ansatz}). 

We now note that the induced 3-metric $g$ takes the block diagonal form
\be 
g = \left( \begin{array}{cc}
U^{-1/3} \, \tilde{g} &  \\
  & U^{2/3}
\label{m2m5.ind.metric}
\end{array} \right) 
\ee
from which it can be seen that the branewave equations for $Y$ reduce to
\be
\pa_{i} \left[ \sqrt{-\det \tilde{g}} \, \tilde{g}^{ij} \, 
\pa_{j} Y^{\mu} \right] = 0               
\ee
where
\be
\tilde{g}_{ij}= \eta_{{\mu}{\nu}} \, \pa_{i}Y^{\mu} \, \pa_{j}Y^{\nu}\, .
\ee
These are the field equations of the Nambu-Goto (NG) action for a string
moving in a D=6 Minkowski spacetime. Thus, the string boundary of the M2-brane
in the M5-brane is governed by the NG string action.

Of course, we could have obtained this result by substituting
(\ref{m2.complete.ansatz}) directly into the $M2$-brane action (\ref{m2action}).
Indeed, it follows from (\ref{m2m5.ind.metric}) that $\det g=\det \gtil$ and
therefore that the $M2$ action collapses to the NG action: 
\be\label{NGact}
S_{M2} \longrightarrow - T \int d^2\sigma \sqrt{-\det \gtil}\, ,
\ee 
where the tension $T$ is given by
\be
T= \left[ \int_0^\infty d \rho \right] \, .
\ee
The tension is infinite because the string is the boundary of an
infinite membrane, but this can be remedied by considering a membrane
stretched between two parallel M5-branes. Let the two M5-branes (with
charges $q$ and $q'$ and worldvolumes aligned with the $Y$ axes) be separated
by a distance $L$ along the $X^1$ axis. This can be achieved by choosing the
harmonic function $U$ to be
\be 
U=1+ {|q|\over |{\bf X}|^3} + {|q'|\over |{\bf X}- \bar{\bf X}|^3}
\ee
where $\bar{\bf X} = (L,0,0,0,0)$. Most of the previous discussion still applies
because the explicit form of the harmonic function $U$ was not used. However,
on the surface $X^2=X^3=X^4=X^5=0$ both $X^1=0$ {\sl and} $X^1=L$ are
now at infinite proper distance, so the membrane coordinate $\rho$ must now
be restricted to take values in the open interval $(0,L)$. In this case 
$T=L$, which is finite\footnote{The $L\rightarrow 0$ limit cannot be taken
because the singularities of $U$ in this limit are genuine curvature
singularities of the M5-brane solution.}.


\section{The 3-brane intersection of two $M5$-branes}

Two M5-branes can have a 1/4 supersymmetric 3-brane intersection. We
shall derive the dynamics of this 3-brane within one of the M5-branes by
replacing the latter by its supergravity solution, given above in terms of the
harmonic function $U$. Let $\xi^I$ ($I=0,1,\dots,5$) be coordinates for
the M5-brane's worldvolume $W$. The M5-brane's Lorentz covariant effective 
action \cite{PST} is   
\bea
S_{M5} &=& \int d \xi^6 \left[ \sqrt{-\det (g + i \tilde{H})} + 
{1\over4}\, \frac{1}{\sqrt{ ( \partial a )^{2}}} \,
\tilde {\cal H}^{IJ} \, H_{IJK} \, \partial^{K} a \right] + \nn
&&+ \int_W \left( {\cal C}_{(6)} + \frac{1}{2} H \wedge  {\cal C}_{(3)}
\right)
\label{m5action}
\eea
where ${\cal C}_{(6)}$ is the pull-back to the worldvolume of the {\sl
on-shell} 6-form dual $C_{(6)}$ of the 3-form potential $C_{(3)}$. The field $a$
is the non-dynamical `PST' gauge field; it can  be eliminated by a choice of
gauge. The worldvolume 3-form $H$ is a `modified' field-strength for a
worldvolume 2-form potential $A$:
\be
H = d A - {\cal C}_{(3)}\, .
\ee
The worldvolume tensor density $\tilde {\cal H}$ is defined by\footnote{This
and the following definition differ slightly from those of \cite{PST}.}
\be
\tilde {\cal H}^{IJ} \equiv  \frac{1}{6\,\sqrt{( \partial a )^{2}}} \, 
\e^{IJKLMN} \, \pa_{K} a \,H_{LMN}
\ee
while the worldvolume 2-form $\tilde H$ has components
\be
\tilde H_{IJ} = {1\over \sqrt{-\det g}}\; g_{IK} \,g_{JL}\, \tilde {\cal H}^{KL}
\ee

We now set $\xi^I = (\sigma^i,\rho,\lambda)$ ($i=0,1,2,3$) and choose a gauge
for which $a= \lambda$. Note that all gauge choices for $a$ appear to break
some symmetry that we wish to keep, but this will not show up in the
final result. Following the previous M2-M5 case, we now seek a vacuum solution 
of the M5-brane's branewave equations that can be interpreted as representing the
intersection on a 3-brane with the fivebrane source of the background. The
appropriate vacuum solution is 
\bea
&& Y^0 =\sigma^0 \sac Y^{1}= \sigma^{1} \sac Y^{2}=\sigma^{2} \sac
Y^{3}=\sigma^{3} \sac Y^4=Y^5=0 \nn
&& X^{1}=\rho \sac X^{2}=\lambda \sac X^3=X^4=X^5=0 \sac H=0
\eea
We shall consider fluctuations about this solution satisfying
\bea
Y^{\mu} &=& Y^{\mu}(\sigma), \nn 
X^{1} &=& \rho, \qquad X^{2} = \lambda, \qquad X^3=X^4=X^5 =0\nn
 i_\rho A &=& {1\over 2} V(\sigma), \qquad i_\lambda A=0, \quad A_{ij}=0
\label{Xansatz}
\eea
where $i_\rho$ and $i_\lambda$ indicate the contraction with the vector fields
$\partial/\partial \rho$ and $\partial/\partial\lambda$, respectively. Note that
the only non-zero component of $H$ is $(i_\rho H)_{ij} = - (dV)_{ij}$.
These conditions constitute a consistent truncation of the full M5-brane
degrees of freedom. An immediate implication is that the induced
worldvolume 6-metric takes the block diagonal form
\be
g = \left( \begin{array}{ccc} 
U^{-1/3} \gtil &  &  \\
& U^{2/3} &  \\
&  & U^{2/3}
\end{array} \right)
\ee
where
\be
\gtil_{ij}= \eta_{\mu\nu} \, \pa_{i}X^{\mu}\pa_{j}X^{\nu}
\ee
Note that $\det g = \det \gtil$. 

A further implication of (\ref{Xansatz}) is that the pull-backs of the 
space-time forms $C_{(3)}$ and $C_{(6)}$ vanish. The worldvolume 3-form ${\cal
C}_{(3)}$ vanishes  for essentially the same reasons as before. To see that
${\cal C}_{(6)}$ also vanishes we recall that it is defined, up to a gauge
transformation, by the relation (see \cite{sorotown} for a review)
\be
d \csix =  \star d \cth - \frac{1}{2} \cth \wedge d \cth \, .
\ee
\newline
In our case this reduces to $d\csix = \star d\cth$ because $\cth \wedge
d \cth$ is a $7$ form on $\bE^5$. One solution of this equation is
\be
\csix=\, U \,dY^0 \wedge \ldots \wedge dY^5\, .
\ee
Any other solution will be a gauge transform of this one, so we may
assume that $\csix$ is of this form. The pullback to the worldvolume of this
form vanishes because it contains (for example) a factor of $\partial
Y^\mu/\partial\rho$, which vanishes for the ansatz (\ref{Xansatz}).

We are nearly ready to extract the 3-brane action. The second term in 
(\ref{m5action}) vanishes upon imposition of (\ref{Xansatz}), so
we just need the 2-form $\tilde{H}$. First note that the only
non-zero components of the tensor density $\tilde{\cal H}$ are 
\be
\tilde{\cal H}^{ij} = \frac 12 \, U^{1/3} \, \e^{ijkl45} \, 
(dV)_{kl}  
\ee
and therefore that the only non-zero components of the 2-form are
\be
\tilde{H}_{ij}= \, U^{-1/3} \, \tilde{B}_{ij} \sac   
\tilde{B}_{ij} \equiv \frac 12 \, \frac{1}{\sqrt{- \det \gtil}} \,
\gtil_{ik} \, \gtil_{jl} \, \e^{klmn} \, (dV)_{mn}
\ee
This immediately implies the following block diagonal form of
the matrix appearing in the first term of the \mf action:
\be
g+i \tilde{H} = \left( \begin{array}{ccc} 
U^{-1/3} \left[ \gtil + i \tilde{B} \right] &  &  \\
& U^{2/3} &  \\
&  & U^{2/3}
\end{array} \right)
\ee
It follows that
\be
\det\left[ g+i \tilde{H} \right] = 
\det \left[ \gtil + i \tilde{B} \right] =
\det \left[ \gtil + dV\right] \, .
\ee
To obtain the last equality one uses firstly that, for any
antisymmetric $4\times 4$ matrix $D$,
\be
\det \, (\gtil + D) = \left(\det \gtil\right) \left[ 1 + \shalf D^{2} + 
\frac{1}{8} (D^{2})^{2} - \frac{1}{4}  D^{4} \right]\, ,
\ee
where $D^{2}=D^{{i}{j}} \, D_{{i}{j}}$ and $D^{4}= D_{{i}{j}}\, D^{{j}{k}}\,
D_{{k}{l}}\, D^{{l}{i}}$, and then that $\tilde{B}^2 = -(dV)^2$
and $\tilde{B}^4=(dV)^4$.

Given these results, the \mf action reduces to 
\be
S_{M5} \longrightarrow T \int d^4 \sigma 
\sqrt{-\det \left[ \gtil + dV \right]}.
\ee
where the tension $T$ is formally infinite, as expected. Apart from this, we
conclude that the dynamics of the 3-brane living in the orthogonal intersection
of two $M5$-branes is governed by the Dirac-Born-Infeld action, at least in the
M5-brane Minkowski vacuum.

\section{Discussion}

In this paper we have used the bosonic sector of the M2-brane and M5-brane
worldvolume actions to derive actions describing the dynamics of M5-brane
intersections corresponding to the 1-brane and 3-brane solitons of the M5-brane
worldvolume field theory. Essentially, we have obtained the latter by a
consistent truncation of the former. In principle, our method could be used to
derive the full supersymmetric action for the 1-brane and 3-brane in the
M5-brane (in a vacuum background) by the simple expedient of retaining fermions
from the beginning. Although we have not done this, we expect that the
resulting actions will be $\kappa$-symmetric extensions of those found here.

The D=6 NG string action is presumably to be interpreted as a special case of a
self-dual string in a more general background that would include a coupling to
the 2-form potential on the M5-brane. It seems likely, in analogy to branes in
spacetime, that $\kappa$-symmetry will require that the background solve the M5
branewave equations. One solution of these equations is $D=6$ Minkowski space
with vanishing 2-form potential, i.e. the M5-brane vacuum. Our method yields the
action for the self-dual string in this vacuum background. The 3-brane action
found here should be similarly interpreted.

The fact that the 3-brane action is of Dirac-Born-Infeld type means that it
has its own worldvolume solitons, which can be interpreted as endpoints of
strings \cite{CM,gib}. It is natural to interpret these strings as the self-dual
strings in the M5-brane. A D=11 spacetime interpretation of this possibility was
given in \cite{bgt}. Thus, the 3-brane is very likely the D-brane of a new
intrinsically non-perturbative self-dual D=6 superstring theory.

\section*{Acknowledgements}
This work in supported in part by NSF grant PHY-9511632, the Robert A.
Welch Foundation AEN95-0590 (CICYT), GRQ93-1047 (CIRIT) and
by the Commission of European Communities CHRX93-0362(04).
D.M. is supported by a fellowship from
Comissionat per a Universitats i Recerca de la Generalitat de
Catalunya.

\end{document}